\documentclass[11pt]{article}
\usepackage{times,amsmath,amsthm,graphicx,amssymb,color,colordvi,subfig}
\usepackage[cm]{fullpage}
\usepackage{wrapfig}
\usepackage{nonfloat}
\usepackage[ruled,vlined,nofillcomment]{algorithm2e}
\usepackage[utf8]{inputenc}
\usepackage{hyperref}

\newcommand{\Degree}{\textit{d}}
\newcommand{\Core}{\textit{k}}

\newcommand{\Index}{\textsl{core}}
\newcommand{\Est}{\textsl{est}}
\newcommand{\Changed}{\textsl{changed}}
\newcommand{\Count}{\textsl{count}}
\newcommand{\Neighbor}{\textsl{neighbor}}
\newcommand{\Again}{\textsl{again}}

\newcommand{\ComputeIndex}{\textsf{\small computeIndex}}
\newcommand{\ImproveEstimate}{\textsf{\small improveEstimate}}

\SetKwFor{ON}{on}{do}{}
\SetKwFor{REPEAT}{repeat}{}{}
\SetKwFor{CLASS}{class}{}{}
\SetKwFor{METHOD}{method}{}{}
\SetKw{AND}{and}
\SetKw{OR}{or}
\SetKw{NOT}{not}
\SetKw{TO}{to}
\SetKw{TRUE}{true}
\SetKw{FALSE}{false}
\SetKw{DOWNTO}{downto}

\newcommand{\INTEGER}{\textbf{int}\xspace}

\newtheorem{theorem}{Theorem}

\newtheorem{corollary}{Corollary}
\newtheorem{definition}{Definition}

\begin{document}

\title{Distributed $k$-Core Decomposition}
\author{Alberto Montresor \\
DISI - University of Trento\\
Via Sommarive 14, I-38123 \\Povo, Trento - Italy\\
{\tt \small alberto.montresor@unitn.it}
\and Francesco De Pellegrini and Daniele Miorandi\\
CREATE-NET\\
Via alla Cascata 56/D, I-38123\\
Povo, Trento -Italy\\
{\tt \small name.surname@create-net.org}
}
\date{}
\maketitle

\begin{abstract}
Among the novel metrics used to study the relative importance of nodes
in complex networks, $k$-core decomposition has found a number of applications in
areas as diverse as sociology, proteinomics, graph visualization, and
distributed system analysis and design. This paper proposes new
distributed algorithms for the computation of the $k$-core decomposition of 
a network, with the purpose of (i) enabling the run-time computation of $k$-cores in ``live'' distributed
systems and (ii) allowing the decomposition, over a set of connected machines, 
of very large graphs, that cannot be hosted in a single machine. Lower bounds on the algorithms
complexity are given, and an exhaustive experimental analysis on real-world
graphs is provided.
\end{abstract}

\section{Introduction}

In the last few years, a number of metrics and methods have been introduced
for studying the relative ``importance'' of nodes within complex network
structures. Examples include betweenness, eigenvector and closeness centrality
indexes~\cite{freeman78,newmanreview}. Such studies have been applied in a variety
of settings, including real networks like the Internet topology, social
networks like co-authorships graphs, protein networks in bio-informatics, and
so on.

\begin{wrapfigure}{r}{0.5\textwidth}
\begin{center}
\includegraphics[width=0.5\textwidth,trim=0 130 0 80]{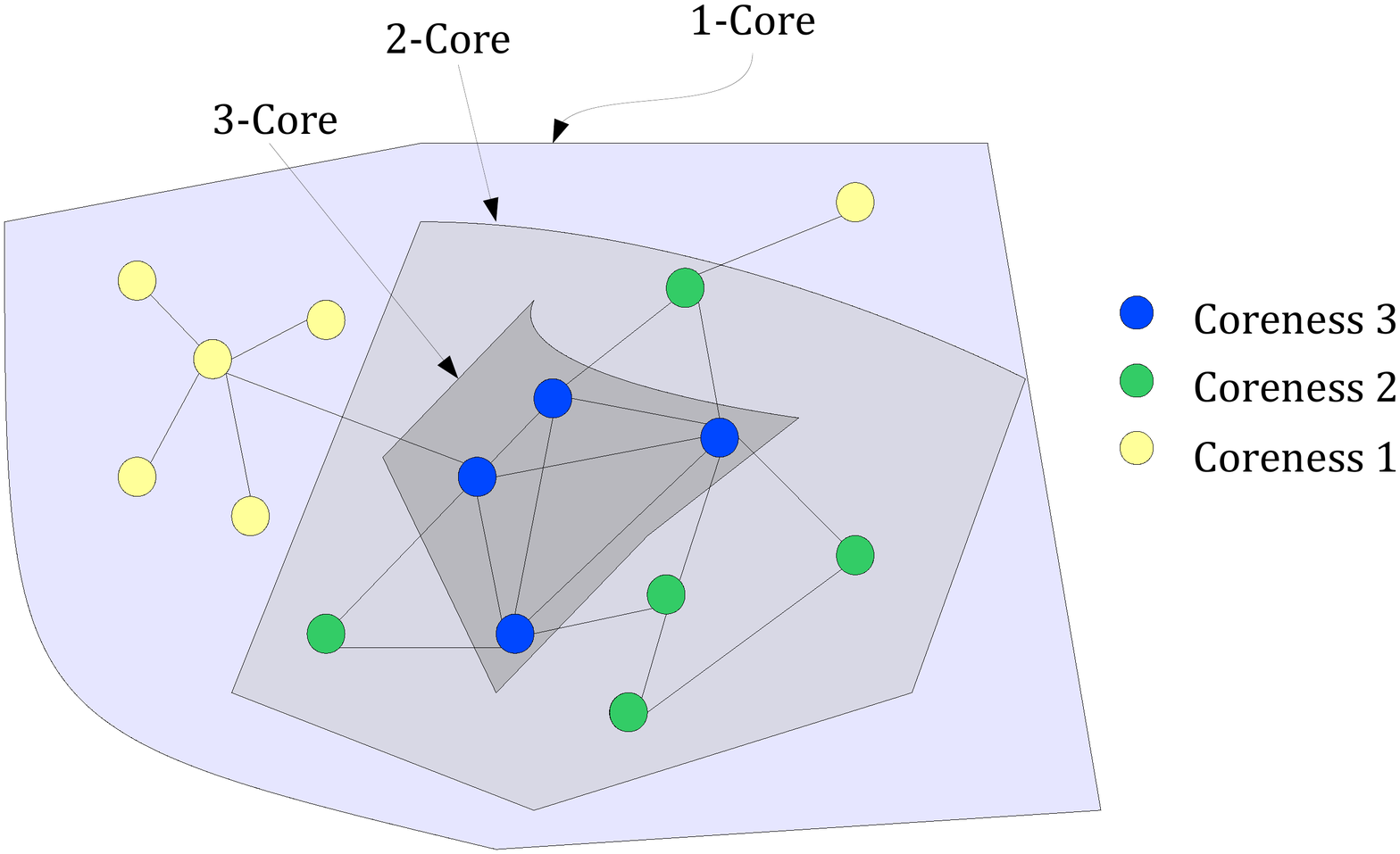}
\end{center}
\caption{$k$–core decomposition for a sample graph.}
\label{f:example-cores}	
\end{wrapfigure}

Among these metrics, $k$-core decomposition is a well-established method for
identifying particular subsets of the graph called $k$-\emph{cores}, or
$k$-\emph{shells}~\cite{seidman1983}. Informally, a $k$-core is obtained by recursively removing all
nodes of degree smaller than $k$, until the degree of all remaining vertices
is larger than or equal to $k$. Nodes are said to have coreness $k$
(or, equivalently, to belong to the $k$-shell) if they belong to the
$k$-core but not to the $(k+1)$-core. As an example of $k$-core decomposition for a sample
graph, consider Figure~\ref{f:example-cores}. Note that by definition cores
are ``concentric'', meaning that nodes belonging to the 3-core belong to the
2-core and 1-core, as well. Larger values of ``coreness'', though, clearly
correspond to nodes with a more central position in the network structure.

$k$-core decomposition has found a number of applications; for example,
it has been used to characterize social networks~\cite{seidman1983},
to help in the visualization of complex graphs~\cite{kcore-visualization06}, 
to determine the role of proteins in complex proteinomic networks~\cite{bader02},
and finally to identify nodes with good ``spreading'' properties in 
epidemiological studies~\cite{kitsak10}.

Centralized algorithms for the $k$-core decomposition already
exist~\cite{batagely03}. Here, we consider the distributed version of this
problem, which is motivated by the following scenarios:
\begin{itemize}
\item \emph{One-to-one scenario}: The graph to be analyzed could be a ``live''
distributed system, such as a P2P overlay, that needs to inspect itself;
\emph{one} host is also \emph{one} node in the graph, and connections among
hosts are the edges. Given that cores with larger $k$ are known to be good
spreaders~\cite{kitsak10}, this information could be used at run-time to
optimize the diffusion of messages in epidemic protocols~\cite{jetstream}.
\item \emph{One-to-many scenario}: The graph could be so large to not fit into
a single host, due to memory restrictions; or its description could be
inherently distributed over a collection of hosts, making it inconvenient to
move each portion to a central site. So, \emph{one} host stores \emph{many}
nodes and their edges. As an example, consider the Facebook social graph, with
500 million users (nodes) and more than 65 billion friend connections (edges)
in December 2010; or the web crawls of Google and Yahoo, which stopped to
announce the size of their indexes in 2005, when they both surpassed the 10 billion
pages (nodes) milestone.
\end{itemize}

Interesting enough, the two scenarios turn out to be related: the former can
be seen as a special case of the ``inherent distribution'' of the latter taken
to its extreme consequences, with each host storing only one node and its
edges.

The contribution of this paper is a novel algorithm that could be adapted to
both scenarios. Reversing the above reasoning, Section~\ref{s:algo} first
proposes a version that can be applied to the one-to-one scenario,
and then shows how to migrate it to the one-to-many scenario, by
efficiently putting a collection of nodes under the responsibility of a single
host. We prove that the resulting algorithm completes the $k$-core
decomposition in $O(N)$ rounds, with $N$ being the number of nodes; more
precisely, Section~\ref{s:proof} shows an upper bound equal to $N-K+1$, with $K$
being the number of nodes with minimal degree, and describes a worst-case
graph that requires exactly such number of rounds. While such upper bounds are
rather high, real world graphs such as the Slashdot comment network, the citation
graph of Arxiv or the Gnutella overlay network require a surprisingly low
number of rounds, as demonstrated in the experiments described in
Section~\ref{s:eval}.

\section{Notation and system model}
\label{s:problem}

Given an undirected graph $G=(V,E)$ with $N=|V|$ nodes and $M=|E|$
edges, 
we define the concept of
\emph{$k$--core decomposition}:

\begin{definition}
A subgraph $G(C)$ induced by the set $C \subseteq V$ is a \emph{k-core} if and
only if $\forall u \in C: \Degree_{G(C)}(u) \geq k$, and $G(C)$ is the maximum
subgraph with this property.
\end{definition}

\begin{definition}
A node in $G$ is said to have \emph{coreness} $k$ if and only if it belongs to
the $k$-core but not the $(k+1)$-core.
\end{definition}

Here, $\Degree_G(u)$ and $\Core_G(u)$ denote the degree and the
coreness of $u$ in $G$, respectively; in what follows, $G$ can be dropped when it is
clear from the context. $G(C) = (C, E|C)$ is the subgraph of $G$ induced by
$C$, where $E|C = \{ (u,v) \in E : u,v \in C \}$.

The distributed system is composed by a collection of hosts $H$, whose overall
goal is to compute the $k$-core decomposition of $G$. Each node
$u$ is associated to exactly one host $h(u) \in H$, that is responsible for
computing the coreness of $u$. Each host $x$ is thus responsible for a
collection of nodes $V(x)$, defined as follows:
\[
  V(x) = \{ u : h(u) = x \}.
\]

Each host $x$ has access to two functions, $\Neighbor_V()$ and $\Neighbor_H()$, that
return a set of \emph{neighbor nodes} and \emph{neighbor hosts}, respectively.
Host $x$ may apply these functions to either itself or to the nodes under its
responsibility; it cannot obtain information about neighbors of other hosts
or nodes under the responsibility of other nodes. Formally, the functions
are defined as follows:
\begin{align*}
\forall u \in V: \Neighbor_V(u) &= \{ v : (u,v) \in E \} \\
\forall x \in H: \Neighbor_V(x) &= \{ v : (u,v) \in E \wedge u \in V(x) \}\\
\forall x \in H: \Neighbor_H(x) &= \{ y : (u,v) \in E \wedge u \in V(x) \wedge v \in V(y) \}
\end{align*}

A special case occurs when the graph to be analyzed coincides with the
distributed system, i.e. $H=V$. When this happens, the label $u$ will be used
to denote both the node and the host, and in general we will use the terms
node and host interchangeably. Also, note that in this case $\Neighbor_V(u)$ =
$\Neighbor_H(u)$.

Hosts communicate through reliable channels. 
For the duration of the computation, we assume 
that hosts do not crash.

\section{Algorithm}
\label{s:algo}

Our distributed algorithm is based on the property of locality of the $k$-core
decomposition: due to the maximality of cores, the coreness of node $u$ is the largest 
value $k$ such that $u$ has at least $k$ neighbors that belong to a $k$-core or a 
larger core. More formally,

\begin{theorem}[Locality]
\label{thm:lemma1}
For each $u \in V$, $\Core(u) = k$ if and only if 
\begin{description}
\item{(i)} there exist a subset
$V_k \subseteq \Neighbor_V(u)$ such that $|V_k| = k$ and $\forall v \in V_k: \Core(v)
\geq k$; 
\item{(ii)} there is no subset $V_{k+1} \subseteq \Neighbor_V(u)$ such that
$|V_{k+1}|=k+1$ and $\forall v \in V_{k+1}: \Core(v) \geq k+1$.
\end{description}
\end{theorem}

\paragraph{Proof} ~
	
\begin{itemize}
\item[$\Rightarrow)$] Since $\Core(u)=k$ there exists a (maximal) set 
$W_k \subseteq V$ such that $u \in W_k$ and $G(W_k)$ is a $k$-core,
and there is no set $W_{k+1} \subseteq V$ 
such that $u \in W_{k+1}$  and $G(W_{k+1})$ is a $(k+1)$-core. Indeed, $G(W_k)$ is a $k$-core for all nodes in 
$W_k$, and so it is for at least $k$ neighbors of $u$, because of maximality. Part (ii) 
follows by contradiction: assume that $v_1, \ldots, v_{k+1}$ are $k+1$ neighbors of $u$ that have coreness $k+1$ or more. 
Denote $W_1, \ldots,W_{k+1}$ the subsets of nodes inducing their corresponding $(k+1)$-cores.
Consider the set $U = \{ u \} \cup \bigcup_{i=1}^{k+1} W_i$, that merges
$u$ with all the $W_i$ sets. The subgraph $G(U)$ induced by $U$ contains at least $k+1$
nodes (because each $W_i$ contains at least $k+1$ nodes); for each node
$v \in U$, $\Degree_{G(U)}(v) \geq k+1$, because $U$ is the union of $k+1$
$(k+1)$-cores and the $k+1$ neighbors of $u$ are included in it.
But this proves that a $(k+1)$-core exists ($G(U)$ may well not be maximal) 
and $u$ belongs to it. Contradiction.

\item[$\Leftarrow)$] For each node $v_i \in V_k$, $1 \leq i \leq k$,
$\Core(v_i)=k$ implies the existence of a set $W_i \subseteq V$ such that
$G(W_i)$ is a $k$-core of $G$ and $v_i \in W_i$. Consider the set $U = \{ u \}
\cup \bigcup_{i=1}^k W_i$. The subgraph $G(U)$ induced by $U$ contains at
least $k+1$ nodes (because each $W_i$ contains at least $k+1$ nodes); for each
node $v \in U$, $\Degree_{G(U)}(v) \geq k+1$, because it is the union of $k$
$k$-cores and the $k$ neighbors of $u$ are included in $U$. Thus, because 
of maximality, $G(U)$ is a $k$-core of $G$ containing $u$. Suppose now that there exists a subset $W
\subseteq V$ such that $G(W)$ is a $(k+1)$-core containing $u$; this means that
$u$ has at least $k+1$ neighbors, each of them with coreness $k+1$ or more;
but this contradicts our hypothesis (ii). We can conclude that $\Core(v) =
k$.\qed
\end{itemize}

The locality property tells us that the information about the coreness of
the neighbors of a node is sufficient to compute its own coreness. Based
on this idea, our algorithm works as follows: each node produces
an \emph{estimate} of its own coreness and communicates it to its neighbors;
at the same time, it receives estimates from its neighbors and use them
to recompute its own estimate; in the case of a change, the new value is
sent to the neighbors and the process goes on until convergence.

This outline must be formalized in a real algorithm; we do it twice, for both
the one-to-one and the one-to-many scenarios. We conclude the section with a
few ideas about termination detection, that are valid for both versions.

\subsection{One host, one node}

Each node $u$ maintains the following variables:
\begin{itemize}
\item $\Index$ is an integer that represents the local estimate of 
the coreness of $u$; it is initialized with the local degree.
\item $\Est[\,]$ is an integer array containing one element for each neighbor;
$\Est[v]$ represents the most up-to-date estimate of the coreness of $v$
known by $u$. In the absence of more precise information, all its entries are
initialized to $+\infty$.
\item $\Changed$ is a Boolean flag set to true if $\Index$ has been
recently modified; initially set to false.
\end{itemize}

The protocol is described in Algorithm~\ref{alg:distr}. Each node
$u$ starts by broadcasting a message $\langle u, \Degree(u) \rangle$ 
containing its identifier and degree to all its neighbors. Whenever $u$
receives a message $\langle v, k \rangle$ such that $k < \Est[v]$, 
the entry $\Est[v]$ is updated with the new value. A new temporary estimate
$t$ is computed by function \ComputeIndex() in
Algorithm~\ref{alg:computeIndex}. If $t$ is smaller than the previously known
value of $\Index$, $\Index$ is modified and the $\Changed$ flag is set to
true. Function \ComputeIndex() returns the largest value $i$ such that there
are at least $i$ entries equal or larger than $i$ in $\Est$, computed
as follows: the first three loops compute how many nodes have estimate
$i$ or more, $1 \leq i \leq k$, and store this value in array $\Count$.
The \textbf{while} loop searches the largest value $i$ such that
$\Count[i] \geq i$, starting from $k$ and going down to $1$.

The protocol execution is divided in periodic rounds: every $\delta$ time
units, variable $\Changed$ is checked; if the local estimate has been
modified, the new value is sent to all the neighbors and $\Changed$ is
set back to false. This periodic behavior is used to avoid flooding
the system with a flow of estimate messages that are immediately superseded
by the following ones.

It is worth remarking that during the execution, variable $\Index$ at
node $u$ (i) is always larger or equal than the real coreness value of $u$,
and (ii) cannot increase upon the receipt of an update message. Informally,
these two observations are the basis of the correctness proof contained
in Section~\ref{s:proof}.

\subsubsection{Example} 

We describe here a run of the algorithm on the simple sample graph reported in
Fig.~\ref{fig:example}. At the first round, all nodes $v$ have
$\Index=\Degree(v)$; nodes $2,3,4$ and $5$ send the same value
$\Index=3$ with their neighbors: these messages do not cause any change in the
estimates of the coreness of receiving nodes. However, in the same
round, nodes $1$ and $6$ notify their $\Index=1$ value to nodes $2$ and $5$,
respectively: as a consequence, node $2$ and $5$ update their estimates to
$\Index=2$. Thus, in the second round another message exchange occurs, since nodes
$2$ and $5$ notify their neighbors that their local estimate changed, i.e.,
they send $\Index=2$ to nodes $1, 3, 4$ and $3, 4, 6$, respectively.
This causes an update $\Index=2$ at nodes $3$ and $4$, which have to send out
another update $\Index=2$ to nodes $2$ and $4$ and nodes $3$ and $5$,
respectively, in the third round. However, no local estimate changes from now on, 
which in turns means that the algorithm converged. Finally, $\Index=2$ for $v=2,3,4,5$ and
$\Index=1$ for $v=1,6$.

\bigskip
\noindent
\begin{minipage}[t]{0.48\linewidth}
\begin{algorithm}[H]
\ON{initialization}{
  $\Index \gets \Degree(u)$\;
  \lForEach{$v \in \Neighbor_V(u)$}{
    $\Est[v] \gets \infty$\;
  }
  \textbf{send} $\langle u, \Index \rangle$ \textbf{to} $\Neighbor_V(u)$\;
}
\BlankLine
\ON{receive $\langle v, k \rangle$}{
  \If{$k < \Est[v]$}{
    $\Est[v] \gets k$\;
    $t \gets \ComputeIndex(\Est, u, \Index)$\;
    \If{$t < \Index$}{
      $\Index \gets t$\;
      $\Changed \gets \mathbf{true}$\;
    }
  }
}
\BlankLine
\REPEAT{every $\delta$ time units (round duration)}{
  \If{$\Changed$}{
  	\textbf{send} $\langle u, \Index \rangle$ \textbf{to} $\Neighbor_V(u)$\;
    $\Changed \gets \mathbf{false}$\;
  }
}
\caption{Distributed algorithm to compute the $k$-core
decomposition, executed by node $u$.}
\label{alg:distr}
\end{algorithm}
\end{minipage}
\hfill
\begin{minipage}{0.48\linewidth}
\begin{algorithm}[H]
\caption{\INTEGER\ \ComputeIndex($\INTEGER{[\,]}\ \Est$, \INTEGER $u$, $k$)}
\lFor{$i=1$ \TO $k$}{
  $\Count[i] \gets 0$\;
}
\ForEach{$v \in \Neighbor_V(u)$}{
  $j \gets \min(k, \Est[v])$\;
  $\Count[j] = \Count[j]+1$\;
}
\For{$i=k$ \DOWNTO $2$}{
  $\Count[i-1] \gets \Count[i-1]+\Count[i]$\;
}
$i \gets k$\;
\While{$i > 1$ \AND $\Count[i] < i$}{
  $i \gets i-1$\;
}
\Return $i$\;

\label{alg:computeIndex}
\end{algorithm}
\begin{center}
\includegraphics[width=\textwidth]{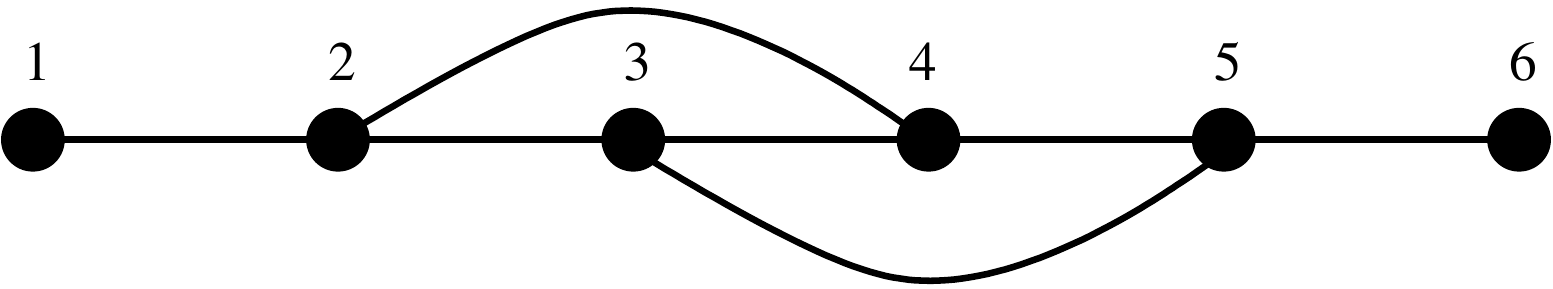}
\end{center}
\figcaption{A simple example describes the run of the algorithm.}
\label{fig:example}
\end{minipage}

\subsubsection{Optimization}\label{s:optim}

Depending on the communication medium available, some optimizations are
possible. For example, if a broadcast medium is used (like in a wireless
network) and the neighbors are all in the broadcast range, the \textbf{send
to} primitive can be actually implemented through a broadcast.
If the \textbf{send to} primitive is implemented through 
point-to-point send operations, a simple optimization is the following:
message updates $\langle u, \Index \rangle$ are sent to a node $v$ if
and only if $\Index < \Est[v]$; in other words, it is sent only
if a node $u$ knows that the new local estimate $\Index$ has the
potential of having an effect on the coreness of $u$; otherwise,
it is skipped. In our experiment, described in Section~\ref{s:eval}, this optimization has shown to be able
to reduce the number of exchanged messages by approximately $50\%$.

\subsection{One host, multiple nodes}

The algorithm described in the previous section can be easily generalized to
the case where a host $x$ is responsible for a collection of nodes $V(x)$: $x$
runs the algorithm on behalf of its nodes, storing the estimates for all of
them and sending messages to the hosts that are responsible for their
neighbors. Described in this way, the new version of the algorithm looks
trivial; an interesting optimization is possible, though. Whenever a host
receives a message for a node $u \in V(x)$, it ``internally emulates'' the
protocol: the estimates received from outside can generate new estimates for
some of the nodes in $V(x)$; in turn, these can generate other estimates,
again in $V(x)$; and so on, until no new internal estimate is generated and
the nodes in $V(x)$ become quiescent. At that point, all the new estimates
that have been produced by this process are sent to the neighboring hosts, where
they can ignite these cascading changes all over again.

Each node $x$ maintains the following variables:
\begin{itemize}
\item $\Est[\,]$ is an integer array containing one element for each node in
$V(x) \cup \Neighbor_V(x)$; $\Est[v]$ represents the most
up-to-date estimate of the coreness of $v$ known by $x$. Given that elements
of $\Neighbor_V(x)$ could belong to $V(x)$ (i.e. some of the
neighbors nodes of nodes in $V(x)$ could be under the responsibility of $x$),
we store all their estimates in $\Est[]$ instead of having a separate array
$\Index[\,]$ for just the nodes in $V(x)$.
\item $\Changed[\,]$ is a Boolean array containing one element for each node
in $V(x)$; $\Changed[v]$ is $\TRUE$ if and only if the estimate of $v$ has
changed since the last broadcast.
\end{itemize}

The protocol is described in Algorithm~\ref{alg:server}. At the beginning, all
nodes $v \in V(X)$ are initialized to $\Est[v] = \Degree(v)$; in the absence
of more precise information, all other entries are initialized to $+\infty$.
Function $\ImproveEstimate()$ is run to compute the best estimates $x$ can
obtain with the local information; then, all the current estimates for the
nodes in $V(x)$ are sent to all nodes.

Whenever a message is received, the array $\Est$ is updated based on the
content of the message; function $\ImproveEstimate()$ is called to take into
account the new information that $x$ may have received.

Periodically, node $x$ computes the set $S$ of all pairs $(v, \Est[v])$ such
that (i) $x$ is responsible for $v$ and (ii) $\Est[v]$ has changed since the
last broadcast. If $S$ is not empty, it is sent to all nodes in the system.

Function $\ImproveEstimate()$ (Algorithm~\ref{alg:improveEstimate}) performs
the local emulation of our algorithm. In the body of the \textbf{while} loop,
$x$ tries and improve the estimates by calling $\ComputeIndex()$ on each of
the nodes it is responsible for. If any of the estimates is changed,
variable $\Again$ is set to \TRUE and the loop is executed another time,
because a variation in the estimate of some node may lead to changes in the
estimate of other nodes.

\subsubsection{Communication policy}

There are two policies for disseminating the estimate updates.
The above version of the algorithm assumes that a broadcast medium
is available. This means that a single message containing all the
updates received since the last round could be created and sent to all.

Alternatively, we could adopt a communication system based on point-to-point
send operations. In this case, it does not make sense to send all updates to
all nodes, because each update is interesting only for a subset of nodes. So,
for each host $y \in H$, we create a message containing only those updates
that could be interesting for $y$. The modification to be applied to
Algorithm~\ref{alg:server} are contained in Algorithm~\ref{alg:modification}.

\subsubsection{Node-hosts assignment policy}

The graph to be analyzed could be ``naturally'' split among the
different hosts, or nodes could be assigned to hosts based on a well-defined
policy. 
It is difficult to 
identify efficient heuristics to perform
the assignment in the general case. In this paper, we adopt a very simple policy: assuming
that nodes identifiers are integers in the range $[0 \ldots n-1]$ and
hosts identifiers are integers in the range $[0 \ldots |H|-1]$, each 
node $u$ is assigned to host $(u \bmod |H|)$.

\subsection{Termination}

To complete both algorithms, we need to discuss a mechanism to detect when
convergence to the correct coreness values has been reached. There are
plentiful of alternatives:
\begin{itemize}
\item {\em Centralized approach}: each host may inform a centralized
server whenever no new estimate is generated during a round; when
all hosts are in this state, messages stop flowing and the protocol
can be terminated. This is particularly suited for the ``one node, multiple hosts''
scenario, where it corresponds to a master-slaves approach.
\item {\em Decentralized approach}: epidemic protocols for
aggregation~\cite{aggregation} enable the decentralized computation of 
global properties in $O(\log |H|)$ rounds. These protocols could be
used to compute the last round in which any of the hosts has generated
a new estimate (namely, the execution time): when this value has not been
updated for a while, hosts may detect the termination of the protocol
and start using the computed coreness.
\item {\em Fixed number of rounds}: as shown in 
Section~\ref{s:eval}, most of real-world graphs can be completed in a very small
number of rounds (few tens); furthermore, after very few rounds the
estimate error is extremely low. If an approximate $k$-core decomposition
could be sufficient, running the protocol for a fixed number of rounds
is an option.
\end{itemize}

\bigskip
\noindent
\begin{minipage}[t]{0.48\linewidth}
\begin{algorithm}[H]
\ON{initialization}{
  \lForEach{$v \in \Neighbor_V(x)$}{
    $\Est[v] \gets +\infty$\;
  }
  \lForEach{$u \in V(x)$}{
	$\Est[u] \gets \Degree(u)$\;
  }
  \ImproveEstimate(\Est)\;
  $S \gets \{ (u, \Est[u]) : u \in V(x) \}$\;
  \textbf{send} $\langle S \rangle$ \textbf{to} $\Neighbor_H(x)$\;
}
\BlankLine
\ON{receive $\langle S \rangle$}{
  \ForEach{$(v, k) \in S$}{
  	\lIf{$k < \Est[v]$}{
      $\Est[v] \gets k$\;
    }
  }
  \ImproveEstimate(\Est)\;
}
\BlankLine
\REPEAT{every $\delta$ time units (round duration)}{
  $S \gets \emptyset$\;
  \ForEach{$u \in V(x)$}{
    \If{$\Changed[u]$}{
  	  $S \gets S \cup \{ (u, \Est[u]) \}$\;
	  $\Changed[u] \gets \FALSE$\;
	}
  }
  \If{$S \neq \emptyset$}{
  	\textbf{send} $\langle S \rangle$ \textbf{to} $\Neighbor_H(x)$\;
  }
}
\caption{Distributed algorithm to compute the $k$-core decomposition, executed by host $x$.}
\label{alg:server}
\end{algorithm}
\end{minipage}
\hfill
\begin{minipage}{0.48\linewidth}
\begin{algorithm}[H]
\caption{\ImproveEstimate($\INTEGER{[\,]}\ \Est$)}	
$\Again \gets \TRUE$\;
\While{$\Again$}{
  $\Again \gets \FALSE$\;
  \ForEach{$u \in V(x)$}{
    $k \gets \ComputeIndex(\Est, u, \Est[u])$\;
    \If{$k < \Est[u]$}{
      $\Est[u] \gets k$\;
	  $\Changed[u] \gets \TRUE$\;
	  $\Again \gets \TRUE$\;
    }
  }
}
\label{alg:improveEstimate}
\end{algorithm}
~\\
\begin{algorithm}[H]
\REPEAT{every $\delta$ time units (round duration)}{
  \ForEach{$y \in \Neighbor_H(x)$}{
    $S \gets \{~ (u, \Est[u]) : u \in V(x) \wedge {}$\\
 	$\qquad\quad (u,v) \in E \wedge v \in V(y)~\}$\;
    \If{$S \neq \emptyset$}{
  	  \textbf{send} $\langle S \rangle$ \textbf{to} $y$\;
    }
  }
  \ForEach{$u \in V(x)$}{$\Changed[u] \gets \FALSE$\;}
}
\caption{Code to be substituted in Algorithm~\ref{alg:server}}
\label{alg:modification}
\end{algorithm}
\end{minipage}

\section{Correctness proofs}
\label{s:proof}

We now prove that our algorithms are correct and eventually terminate. While
we focus on the one-to-one scenario, the results can be easily
extended to the one-to-many case.

\subsection{Safety and liveness}

\begin{theorem}[Safety]
\label{thm:safety}
During the execution, variable $\Index$ at each node $u$ is always larger or
equal than $\Core(u)$.
\end{theorem}

\noindent \textbf{Proof}.
By contradiction, suppose there exists a node $u_1$ such that $\Index(u_1) <
\Core(u_1)$. By Theorem~\ref{thm:lemma1}, there is a set $V_1$ such that
$|V_1|=\Core(u)$ and for each $v \in V_1 : \Core(v) \geq \Core(u)$. In order to set
$\Index(u_1)$ smaller than $k(u)$, $u_1$ must have received a message containing
an estimate smaller than $k(u)$ from at least one of the nodes in $V_1$. Formally, $u_1$
received a message $\langle u_2, \Index(u_2) \rangle$ from $u_2$ at time
$t_2$, such that $u_2 \in V_1$ and $\Index(u_2) < \Core(u_1)$. Given that
$\Core(u_1) \leq \Core(u_2)$ (because $u_2 \in V_1$), we conclude that
$\Index(u_2) < \Core(u_2)$: in other words, we found another node whose
estimate is smaller than its coreness. By applying Theorem~\ref{thm:lemma1}
again, we derive that $u_2$ received a message $\langle u_2, \Index(u_3)
\rangle$ from $u_3$ at time $t_3 < t_2$, such that $\Index(u_3) < \Core(u_2)
\leq \Core(u_3)$. This reasoning leads to an infinite sequence of nodes $u_1,
u_2, u_3, \ldots$ such that $\Index(u_i) < \Core(u_i)$ and $u_i$ received a
message from $u_{i+1}$ at time $t_i$, with $t_i > t_{i+1}$. Given the finite
number of nodes, this sequence contains a cycle $u_i, u_{i+1}, \ldots, u_j =
u_i$; but this means $t_i > t_{i+1} > \ldots > t_j = t_i$, a contradiction.
\qed

\begin{theorem}[Liveness]\label{thm:liveness}
There is a time after which the variable $\Index$ at each node $u$ is always
equal to $\Core(u)$.
\end{theorem}

\noindent \textbf{Proof}.
By Theorem~\ref{thm:safety} variable $\Index(u)$ cannot be smaller than
$\Core(u)$; by construction, variable $\Index$ cannot grow. So, if we prove
that the estimate will eventually become equal to the actual coreness, we have
proven the theorem. The proof is by induction on the coreness $\Core(u)$.
\begin{itemize}
\item $\Core(u)=0$; in this case, $u$ is isolated. Its degree, used to
initialize $\Index$, is equal to its coreness and the protocol terminates at
the very beginning.
\item $\Core(u)=1$; by contradiction, assume that $\Index(u)>1$ never
converges to $\Core(u)$. This means that there is at least one node $u_1$ with
coreness $\Core(u_1)=1$, neighbor of $u=u_0$, that will
never send a message $\langle u_1, 1 \rangle$ to $u_0$ because its variable
$\Index$ never reaches $1$ ($\Index(u_1) > 1$). Reasoning in the same way, we can find
another node $u_2$, different from $u_0$, such that $(u_1, u_2) \in E$, $\Core(u_2)=1$ and
$\Index(u_2) > 1$ forever. Going on in this way, we can build an infinite
sequence of nodes $u_0, u_1, u_2, \ldots$ connected to each other, all of them
having $\Core(u_i)=1$ and $\Index(u_i)>1$ and such that $u_i \neq u_{i-2}$ for
$i \geq 2$. Given the finite number of nodes, there is at least one cycle with three nodes or more 
in this sequence;  but all nodes belonging to such a
cycle would have coreness at least $2$, a contradiction.
\item \emph{Induction step}: by contradiction, suppose there is a node $u_1$
such that $\Core(u_1)=k>1$ and $\Index(u_1)>k$ forever. By
Theorem~\ref{thm:lemma1}, there are $f \geq k$ neighbors of $u$ with coreness
greater or equal than $k$, and $\Degree(u)-f$ neighbors of $u$ whose coreness
is smaller than $k$. If $f=k$, $u_1$ will eventually receive $\Degree(u_1)-k$
estimates smaller than $k$ (by induction), while the other $k$ estimates will
always be larger or equal to $k$ (by Theorem~\ref{thm:safety}). So, $u_1$
eventually sets $\Index(u_1)$ equal to $k$, a contradiction. If $f \geq k+1$,
there is at least one node $u_2$ among those $f$ such that $\Core(u_2)=k$
(otherwise, having $f \geq k+1$ neighbors with coreness $k+1$ or more,
$\Core(u_1)$ would be $k+1$ or more, a contradiction) and $\Index(u_2)>k$
forever (otherwise, $u_1$ would have received $f \geq k+1$ updates equal to
$k$ or more, setting $\Index(u_1) = k$, a contradiction). Note that the
remaining $f-1 \geq k$ neighbors of $u_1$ have coreness $k+1$ or more. By
reasoning similarly as above, we can build an infinite sequence of nodes $u_1,
u_2, u_3, \ldots$ such that $\Core(u_i)=k$, $\Degree(u_i)>k$ and $\Index(u_i)>k$,
with $u_i$ is waiting a message from $u_{i+1}$ to lower $\Index(u_i)$ to $k$.
As above, the finite number of nodes implies that the sequence contains at
least one cycle $C=\{u_i, u_{i+1}, u_{i+2}, \ldots, u_j = u_i$. Now, for each
of the nodes $u_i \in C$, consider $k$ neighbors $v_{i}^1 \ldots v_{i}^{k}$ of
$u_i$ such that $\Core(v_i^j)>k$. Let $V_i^j$ be a $(k+1)$-core containing
$v_i^j$ (such cores exist because their coreness is larger than $k$). Consider
now the set $U$ defined as the union of all nodes $u_i \in C$ and all
$(k+1)$-cores defined above:
\[
  U = C \cup \bigcup_{u_i \in C, 1 \leq j \leq k} V_i^j
\]
Consider the subgraph $G(U)$ induced by $U$; in such graph, all nodes have at
least $k+1$ neighbors, because the nodes in the cores $V_i^j$ have at least
$k+1$ neighbors and each the nodes in $C$ have $k$ neighbors plus a distinct
node that follows in the cycle (by construction). Thus, $G(U)$ is a
$(k+1)$-core containing $C$, contradicting the assumption that the nodes in
$C$ have coreness $k$.
\qed
\end{itemize}

\subsection{Time complexity}

We proved that our algorithms eventually converge to the correct coreness; we
now discuss upper bounds on the \emph{execution time}, defined as the total
number of rounds during which at least one node broadcasts its new estimate
(when no new estimates are produced, the algorithm stops and the correct
values have been obtained).

For this purpose, we assume that rounds are synchronous; during one round, each
node receives all messages addressed to it in the previous round (if any),
computes a new coreness estimate and broadcasts a message to all its neighbors
if the estimate has changed with respect to the previous round. At round $1$, each node
broadcasts its current estimate (equal to its degree) to all its neighbors. To
simplify the analysis, no further optimizations are applied. In the final round,
messages are sent but they do not cause any variation in the estimates, so
the protocol terminates.

The first observation is that after the first round, in any subsequent round
before the final one at least one node must change its own estimate, reducing
it by at least $1$. This brings to the following theorem:

\begin{theorem}\label{thm:b1}
Given a graph $G=(V,E)$, the execution time is bounded by
$1 + \sum\limits_{u \in V} \left[\Degree(u)-\Core(u)\right]$.
\end{theorem}

\noindent \textbf{Proof}.
The quantity $\left[\Degree(u)-\Core(u) \right]$ represents the
``initial error'' at node $u$,
i.e. the difference between the initial estimate (the degree) and the actual
coreness of $u$. In the worst case, at most one message is broadcast per round, and
each broadcast reduces the error by one unit, apart from the last one which
has no effect. Thus the execution time is bounded by the sum of all
initial errors plus one.\qed

While the previous bound is based on the knowledge of the actual
coreness index of nodes, we can define a bound on the execution time that depends only on
the graph size:

\begin{theorem}\label{thm:b2}
The execution time is not larger than $N$.
\end{theorem}

\noindent \textbf{Proof}.
Given a run of the algorithm, denote 
$
  A(r) = \{ u \in V : \Index(u)=\Core(u) \;\mbox{at round $r$ } \}.
$
We make the following observations: 
\begin{itemize}
\item[i)] $A(1) \neq \emptyset$: each node $u$ with minimal degree $\delta$ is
included in $A(1)$. In fact, $u$ is such that $\Core(u) = \delta$, otherwise
there would be a node $v \in \Neighbor_V(u)$ with a degree less than $\delta$,
which is impossible because $\delta$ is minimal. Given that $\Index(u)=\delta$
at round $1$ by initialization, $u$ belongs to $A(1)$.
\item[ii)] If $u \in A(r)$, then $u$ does not send any message for all
remaining rounds $r+2,r+3,\ldots$.
\item[iii)] $A(r)\subseteq A(r+1)$ $\forall r$.
\end{itemize}
We denote by $T$ the smallest round index at which $A(T)=V$. By
definition, the execution time equals $T+1$.\footnote{This is due to
  the fact that, by our definition, the execution time includes also the
  last round, in which updates are sent but they have no further
  effect on the computed coreness.}

Denote $m(r) = \min \{ \Core(u) : u \not \in A(r) \}$, i.e., the minimal
coreness of a node that did not yet attain the correct value at round $r$.
Also, denote $M(r)=\{ v : \Core(v)=m(r), v \not\in A(r) \}$, the set of all
such nodes.

Assume $A(r) \neq V$ so that $M(r) \neq \emptyset$: at round $r+1$ there must
exist $v \in M(r)$ such that $v\in A(r+1)$, i.e., $v$ attains the correct
value at round $r+1$. In fact, observe that at rounds $r+2,r+3,\ldots$, only nodes in
$M(r)$ can exchange $m(r)$ values due to ii). Thus, if no node in $M(r)$ has
attained the correct value at round  $r+1$, it means that all nodes in $M(r)$ have 
at least $m(r)+1$ neighbors whose estimates is larger than $m(r)$ at round $r+1$.
However, nodes with $\Core(v) \leq m(r)$ that belong to $A(r)$ will never
notify such value again. But, by definition of $m(r)$, no lesser estimate  
will be broadcast. Hence, the correct estimate $m(r)$ at such nodes will
never be attained, contradicting Theorem~\ref{thm:liveness}.

We hence proved that $D(r)=A(r) \setminus A(r-1) \neq \emptyset$ for
$r=1,\ldots,T$, where we let $A(0)=\emptyset$ for the sake of notation and 
$A(1)\not=\emptyset$ because of i). Also, it is easy to see that 
$V=A(T)=\cup_{r=1}^T D(r)$ and $D(r) \cap D(s)=\emptyset$ for $r\not  =s$. Thus,
\[
N = |\bigcup\limits_{r=1}^T D(r)|= \sum_{r=1}^T |D(r)| \geq T
\]

The tighter bound $T \leq N - 1$ is obtained by contradiction. Consider round
$N-2$ and assume $T>N-1$. Using the same arguments as above, $|A(N-2)|\geq N-2$.

Case $|A(N-2)|=N-1$: the only remaining node $v$ such that $\Index(v) \neq \Core(v)$ 
would obtain the true coreness of its neighbors at round $N-1$, against our assumption.

Case $|A(N-2)|=N-2$: let us denote $v_{1}$ and $v_{2}$ the pair of nodes such that
$\Index(v_i) \neq \Core(v_i)$ at round $N-2$. It is easy to see that such
nodes must be neighbors, otherwise all their neighbors would have the correct
$\Index$ value and they would receive those estimates and computed the
correct value by round $N-1$. Also, they both have all remaining neighbors in the set
$A(N-2)$, otherwise one of them would have degree $1$, which is not possible 
since it would belong to $A(1)$. 
However, $\Index(v_i) = \Core(v_i)+1$ for $i=1,2$: in fact 
$\Index(v_i)>\Core(v_i)$ at round $N-1$ and exactly one neighbor 
has a wrong estimation. Also, $\Index(v_1)\geq k_{v_2}+1$ and 
$\Index(v_2) \geq k_{v_1}+1$. Thus, $\Index(v_1) = k_{v_1}+1\geq
k_{v_2}+1$ and also $\Index(v_2) = k_{v_2}+1\geq k_{v_1}+1$ so that
$\Index(v_1) = \Index(v_2)$. However, nodes in $A(N-2)$ will 
not notify again their correct estimate from round $N$ on and  
nodes $v_1$ and $v_2$ will perform the same estimate they had at 
round $N-1$, i.e., $k_{v_2}+1=\Index(v_1) = \Index(v_2)=k_{v_1}+1$. 
Thus, no message can be exchanged from round $N$ on, while $\Index(v_i)\not =\Core(v_i)$ $i=1,2$. 
But, this contradicts the liveness property so that it must be $T \leq N - 1$. 
\qed


From the proof, we observe that the nodes of minimal degree attain the correct
coreness at the first round. We can slightly refine the bound as:

\begin{corollary}\label{thm:cor_b2}
Let $K$ be the number of nodes with minimal degree in $G$. Then the execution
time on $G$ is not larger than $N-K+1$ rounds.
\end{corollary}

It is worth remarking that the bound provided by Theorem~\ref{thm:b2}
is tighter than that provided by Theorem~\ref{thm:b1} if and only if
the initial average estimation error $\frac{ \sum\limits_{u \in V}
  \Degree(u) - \Core(u)}{N}$ is larger than $1-\frac{1}{N}$.

Some important questions are (i) how tight is the bound of Theorem~\ref{thm:b2}, and (ii) is there
any graph that actually requires $N$ rounds to complete? Experimental
results with real-life graphs show that the bound is far from being tight (graphs
with millions of nodes converge in less than one hundred rounds). 
However, we managed to identify a class of graphs close to the bound,
i.e., with execution time equal to $N-1$ rounds for $N\geq 5$.
Assuming that nodes are numbered from $1$ to $N$, the rules to build such
graphs are:
\begin{itemize}
\setlength{\itemsep}{0pt}
\item node $N$ is connected to all nodes  apart from node $N-3$;
\item each node $i=1 \ldots N-2$ is connected with its successor $i+1$;
\item node $N-3$ is also connected with node $N-1$.
\end{itemize}

\begin{figure}[t]
\begin{center}
\includegraphics[width=40mm]{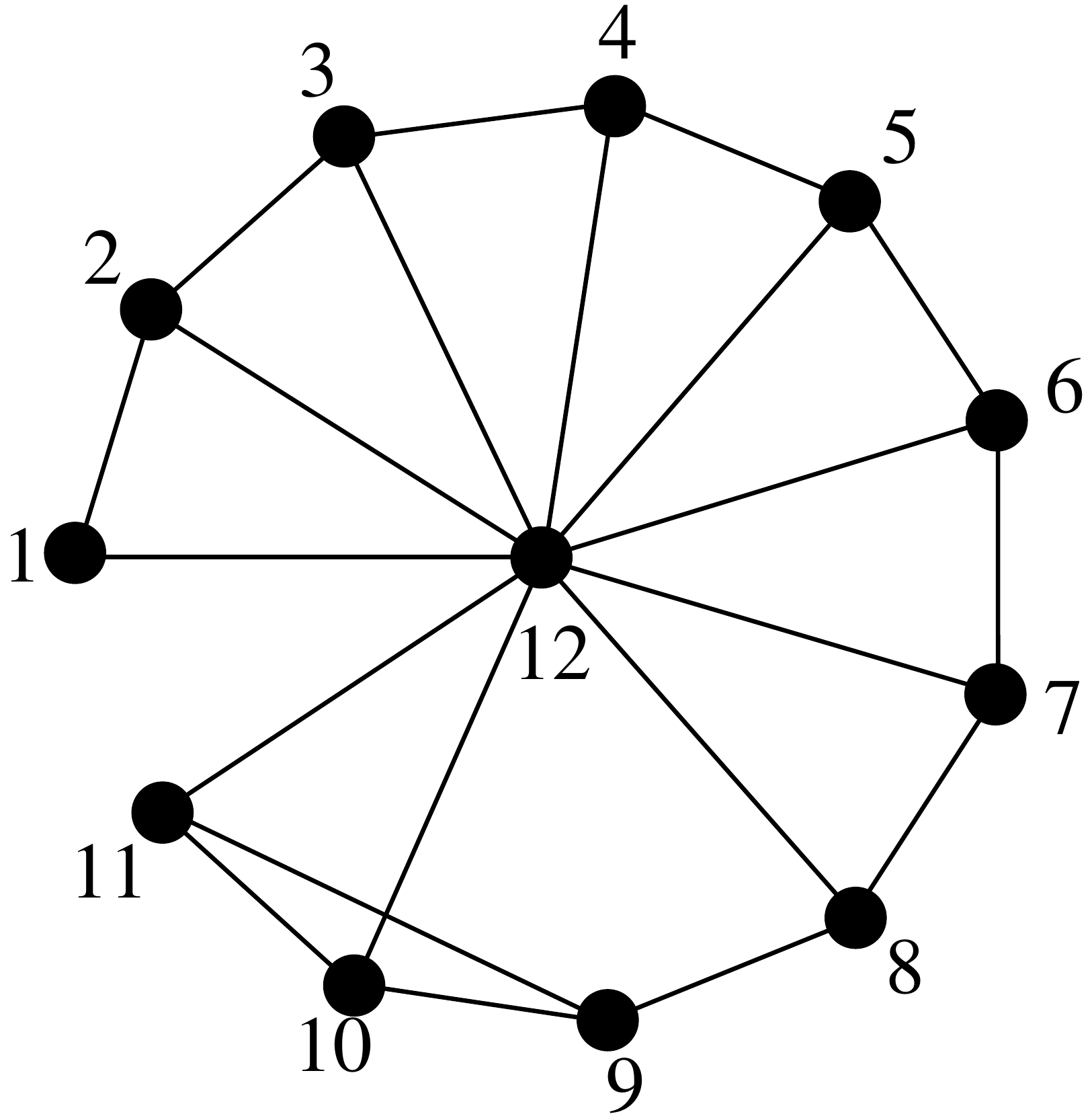}
\end{center}
\vspace{-12pt}
\caption{The worst-case graph, for which the execution time is exactly
  $N-1$ rounds, $N=12$.}
\label{fig:long_rounds}
\end{figure}

Figure~\ref{fig:long_rounds} shows the graph obtained by this scheme for
$N=12$. Graphically, it is convenient to represent node $N$ as the \emph{hub}
of a polygon, where nodes are located at the corners. All nodes have degree
$3$, apart from the hub which has degree $N-2$ and node $1$ which has degree
$2$.
When starting our algorithm, node $1$ acts as a \emph{trigger}: it has the
smallest degree and its broadcast causes node $2$ to change its estimate to
$2$, which in turn will cause node $3$ to change its estimate to $2$, and so
on until the estimate of node $N-4$ changes to $2$. Note that node $N$ has
changed its estimate from $N-2$ to $3$ after the first round, and has
maintained this estimate so far. In the next next round, nodes $N-3$ and $N$
change their estimate to $2$; in the last round, node $N-1$ and $N-2$ change
their estimate to $2$ as well and the algorithm terminates. Given that during
each round apart from the last two, at most one node has changed its estimate,
the total number of rounds is exactly $N-1$ ($N-2$ plus the last round).

It is worth remarking that other simple structures one may think of as
potential worst cases offer lower execution time. As an example, a linear chain of size $N$ requires $\lceil N/2 \rceil$
rounds to converge.

One would expect that there there should be a relation between diameter and
execution time. 
The smaller the diameter, the shorter
should be the execution time. However, despite we noticed a beneficial effect
of small diameters, this does not hold in general: in fact, the example of
Figure~\ref{fig:long_rounds} provides a case when the convergence time
increases linearly with $N$ but the diameter is $3$, i.e., a
constant regardless of $N$.

\subsection{Message complexity}

The maximum number of exchanged messages can be computed
using a double counting argument: during the run of the algorithm, each node
$u$ can at most receive $\Degree(v)-\Core(v)$ updates from each neighbor $v
\in \Neighbor_V(u)$. Then, there are at most $\Degree(u)+\Degree(v)-2$ messages
that can be exchanged over link $uv$. 
If we sum over all the links
\begin{equation}\label{eq:msgcomplx}
\sum_{(u,v) \in E} \left[\Degree(u)+\Degree(v) -2 \right] =\left[ \sum_{v \in
  V(G)} \Degree^2(v) \right] -2\cdot M \leq 2 M \left(\Delta -1\right)
\end{equation}
where $\Delta$ is the maximum degree in the graph. Overall, we obtain the
following worst case bound:
\begin{corollary} Give a graph $G=(V,E)$, the
message complexity is bounded by $\left[\sum_{v \in V(G)}
  \Degree^2(v)\right]- 2M$. 
\end{corollary}
Looking at the left hand-side of (\ref{eq:msgcomplx}) we can see that
the message complexity of the distributed $k$-core
computation is $O(\Delta\cdot M)$.

\section{Experimental evaluation}
\label{s:eval}

This section reports experimental results for both the one-to-one and the
one-to-many versions of the algorithm, over a selection of graphs contained in
the Stanford Large Network Dataset
collection~\footnote{\texttt{http://snap.stanford.edu/data/}}. Undirected
graphs have been transformed in directed graphs by considering both
directions (i.e., two
edges) for each link present in the original one.

Simulations have been performed using Peersim~\cite{peersim}. 
Time is still measured in rounds, i.e. fixed-size time
intervals during which each node has the opportunity to send one update
message to all its neighbors. Unless otherwise stated, the results show the
average over $50$ experiments. Experiments differ in the (random)
order with which operations performed at different
nodes are considered in the simulation. 

\subsection{One-to-one version}

\begin{table*}
\begin{center}\footnotesize
\begin{tabular}{|l|r|r|r|r|r|r||r|r|r|r|r|}
\hline
Name & $|V|$ & $|E|$ & $\oslash$ & $d_{\mathit{max}}$ & $k_{\mathit{max}}$& $k_{\mathit{avg}}$ & $t_{\mathit{avg}}$ & $t_{\mathit{min}}$ & $t_{\mathit{max}}$ & $m_{\mathit{avg}}$ & $m_{\mathit{max}}$\\
\hline
1) CA-AstroPh              &     18\,772 &    198\,110 &  14 &        504 &  56 & 12.62 &  19.55 &  18 &  21 & 47.21 &      807.05\\
\hline
2) CA-CondMat              &     23\,133 &     93\,497 &  15 &      280 &  25 &  4.90 &  15.65 &  14 &  17 & 13.97 &      410.25\\
\hline
3) p2p-Gnutella31          &     62\,590 &    147\,895 &  11 &       95 &   6 &  2.52 &  27.45 &  25 &  30 &  9.30 &      131.25\\
\hline
4) soc-sign-Slashdot090221 &     82\,145 &    500\,485 &  11 &   2\,553 &  54 &  6.22 &  25.10 &  24 &  26 & 29.32 &   3\,192.40\\
\hline
5) soc-Slashdot0902        &     82\,173 &    582\,537 &  12 &   2\,548 &  56 &  7.22 &  21.15 &  20 &  22 & 31.35 &   3\,319.95\\
\hline
6) Amazon0601              &    403\,399 & 2\,443\,412 &  21 &   2\,752 &  10 &  7.22 &  55.65 &  53 &  59 & 24.91 &   2\,900.30\\
\hline
7) web-BerkStan            &    685\,235 & 6\,649\,474 & 669 &  84\,230 & 201 & 11.11 & 306.15 & 294 & 322 & 29.04 &  86\,293.20\\
\hline
8) roadNet-TX              & 1\,379\,922 & 1\,921\,664 & 1049&       12 &   3 &  1.79 &  98.60 &  94 & 103 &  4.45 &       19.30\\
\hline
9) wiki-Talk               & 2\,394\,390 & 4\,659\,569 &   9 & 100\,029 & 131 &  1.96 &  31.60 &  30 &  33 &  5.89 & 103\,895.35\\
\hline
\end{tabular}
\end{center}
\caption{Results with the one-to-one algorithm. Name of the data set,
number of nodes, number of edges, diameter, maximum degree, maximum coreness, average coreness, average-minimum-maximum number of cycles to
complete, average/maximum number of messages sent per node.}
\label{fig:table}
\end{table*}

For this version, the main results are summarized in Table~\ref{fig:table},
which is divided in two parts. On the left, the main features of each
graph considered are reported: name, number of nodes, number of edges, diameter, maximum
degree, to conclude with maximum and average coreness.

On the right, the table reports information about the performance of the
one-to-one protocol, based on two figures of merit: execution time (measured
as the number of rounds in which at least one node sends an update message)
and total number of messages exchanged. In particular, $t_{\mathit{avg}}$,
$t_{\mathit{min}}$ and $t_{\mathit{max}}$ represent the average, minimum and
maximum execution time measured over $50$ experiments. $m_{avg}$ and $m_{max}$
represent the average and maximum number of messages per node.

A few observations are in order. First of all, the execution time is of the
order of few tens of rounds for most of the graphs, with only a couple of them
requiring few hundreds of rounds (web-Berkstan, the web graph of Berkeley and
Stanford, and RoadNet-TX, the road network of Texas). Compared with our
theoretical upper bounds (number of nodes and total initial error),
this suggests 
that our algorithm can be efficiently used in real-world settings.

The average and maximum number of messages per node is, in general, comparable
to the average and maximum degree of nodes. Clearly, nodes with several
thousands neighbors will be more overloaded than others.

In order to understand why web-Berkstan requires so many rounds to
complete, we performed an in-depth analysis of the dynamic behaviour
of the proposed algorithms. In particular, we considered, for each
core, the time taken for all nodes within it to reach the correct
coreness value. Results are reported in
Table~\ref{table:berkstan}. The first two columns report the
problematic cores and their cardinality, respectively. 
The remaining columns represent
the percentage of nodes whose estimate is still erroneous at round $t=25,50, \ldots,
300$; an empty column
corresponds to $0$\%, i.e. the core computation has been completed. At first
look, the $55$-core seems particularly problematic, given that more than one
half of it is still incorrect at round $25$. But the $55$-core completes
before round $225$, well before the $1$-core that terminates after round $300$.
Delays in computing the $1$-core may be associated to the high diameter of
this particular graph, with ``deep'' pages very far away from the highest
cores.

\begin{table*}
\begin{center}\footnotesize
\begin{tabular}{|r||r|r|r|r|r|r|r|r|r|r|r|r|r|}
\hline
k   &  \#     & 25 & 50 & 75 & 100 & 125 & 150 & 175 & 200 & 225 & 250 & 275 & 300 \\
\hline
\hline
 1 & 55\ 776& 14.12\% & 10.26\% & 7.36\% & 4.97\% & 2.99\% & 1.65\% & 0.92\% & 0.56\% & 0.21\% & 0.13\% & 0.08\% & 0.02\% \\
\hline
 2 & 83\ 109 & 3.81\% & 1.35\% & 0.55\% & 0.27\% & 0.14\% & 0.06\% &  &  &  &  &  &  \\
\hline
 3 & 67\ 910 & 1.42\% & 0.23\% &  &  &  &  &  &  &  &  &  &  \\
\hline
 4 & 44\ 548 & 0.95\% & 0.07\% &  &  &  &  &  &  &  &  &  &  \\
\hline
 5 & 68\ 728 & 0.46\% & 0.05\% &  &  &  &  &  &  &  &  &  &  \\
\hline
 6 & 35\ 985 & 3.48\% & 1.01\% & 0.01\% &  &  &  &  &  &  &  &  &  \\
\hline
 8 & 32\ 412 & 1.21\% & 0.46\% & 0.10\% &  &  &  &  &  &  &  &  &  \\ 
\hline
 9 & 28\ 042 & 0.18\% &  &  &  &  &  &  &  &  &  &  &  \\
\hline
10 & 22\ 322 & 1.96\% & 0.64\% &  &  &  &  &  &  &  &  &  &  \\
\hline
15 &  6\ 842 & 0.99\% &  &  &  &  &  &  &  &  &  &  &  \\
\hline
55 &  2\ 548 & 50.78\% & 43.84\% & 36.77\% & 29.71\% & 22.76\% & 15.46\% & 8.40\% & 1.73\% &  &  &  &  \\
\hline
\end{tabular}
\end{center}
\caption{{\small Information about nodes that are delaying the completion of the
protocol in the web-Berkstan graph. The first column $k$ represents a coreness
value; the second column \# represents the size of the $k$-core, i.e. the
number of nodes whose coreness is $k$; the column labeled $t=25, 50, \ldots,
300$ represents the percentage of nodes in the given core that do not know the
correct coreness value after $t$ rounds. Empty cells corresponds to $0$\%. All
other coreness are correctly computed at round $25$.}}
\label{table:berkstan}
\end{table*}

\begin{figure}[t]
\begin{center}
\includegraphics[width=9cm]{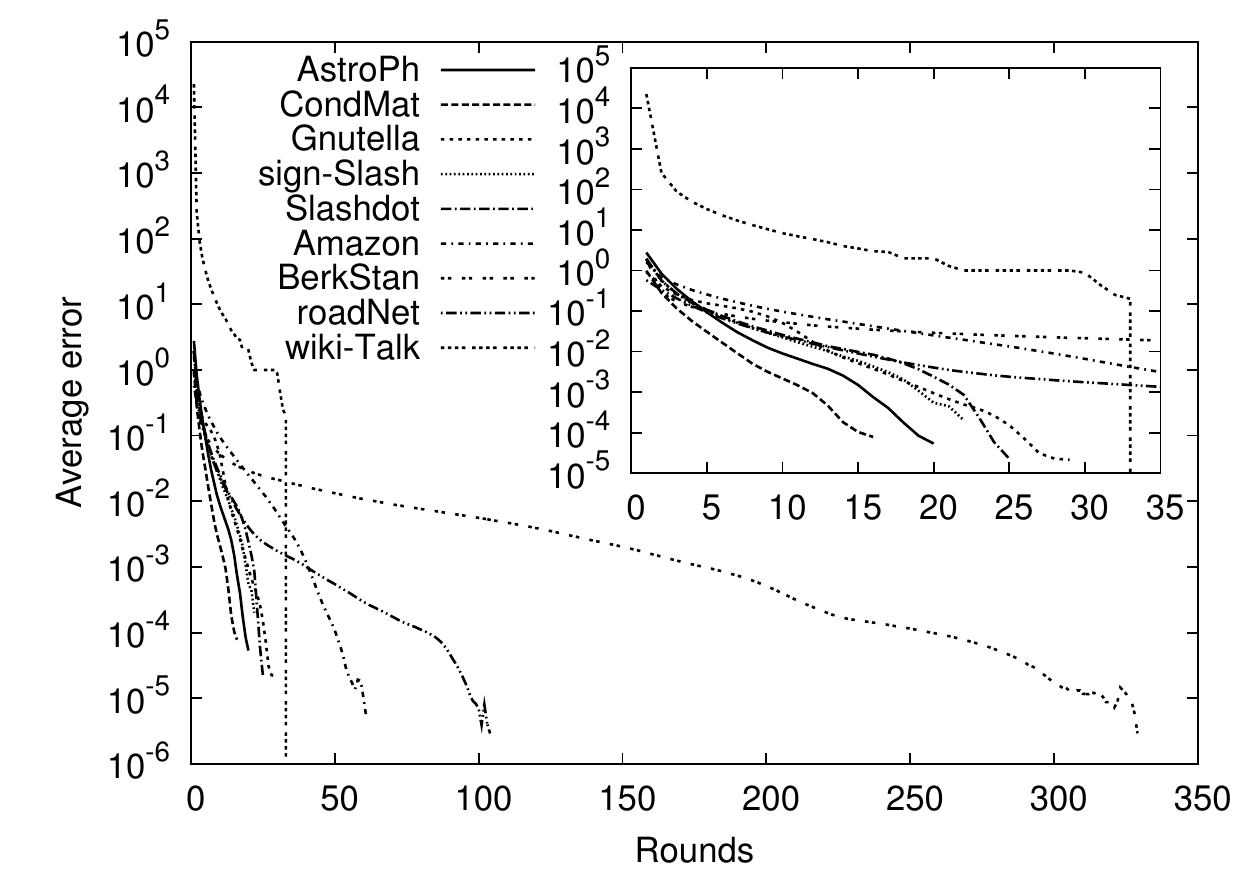}
\includegraphics[width=9cm]{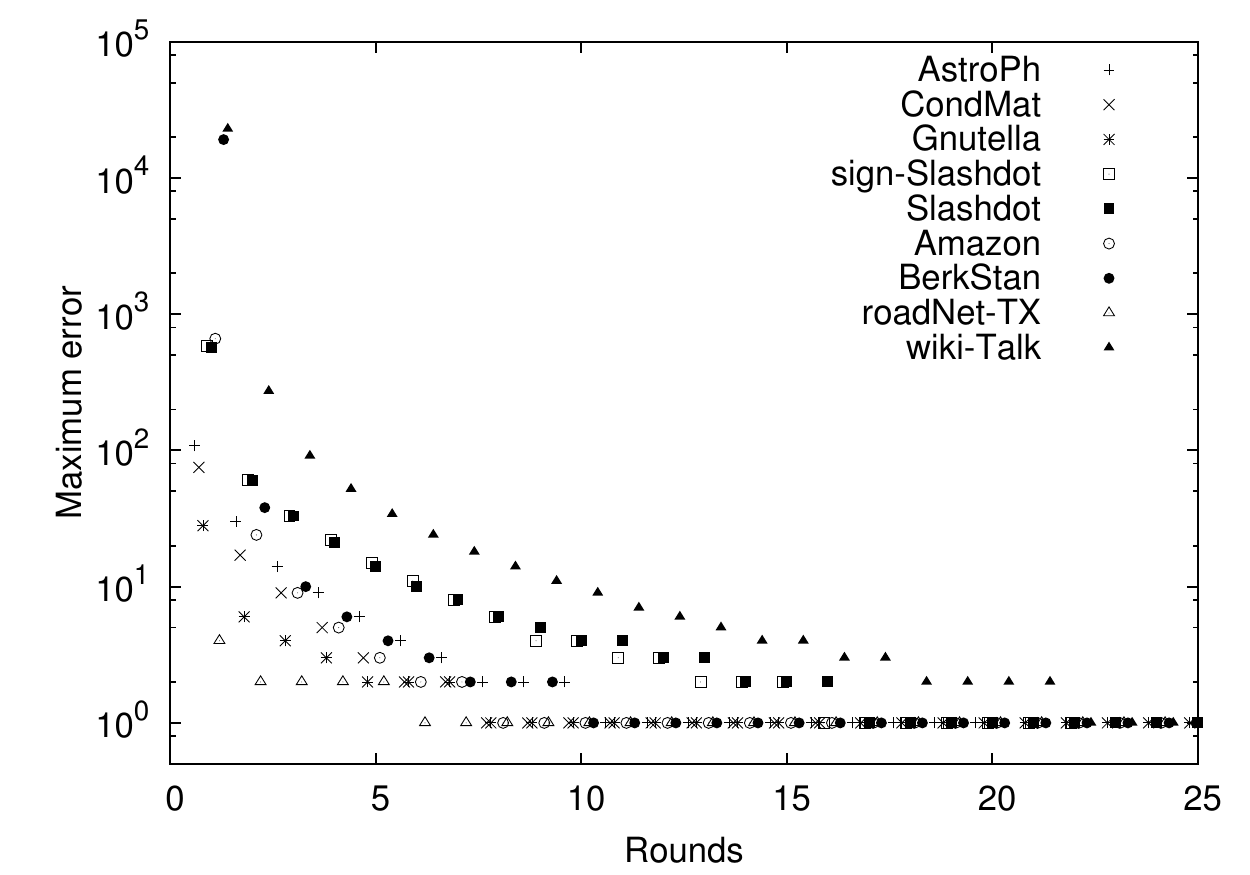}
\end{center}
\caption{Evolution of evaluation error over time. On the left, the average
error over all nodes and all repetitions is shown. The smaller graph shows the details of
the first rounds of the computation. The right part shows the maximum error over
all nodes and all repetitions.}
\label{fig:plot-avg-max}
\end{figure}

Another figure of merit is the temporal evolution of \emph{error}, measured as
the difference -- at each node -- between the current estimate of the coreness
and its correct value. The left part of Figure~\ref{fig:plot-avg-max} shows the
average error for our experimental graphs. When the line stops, it means that
the algorithm has reached the correct coreness estimate, so the error
is zero. 
The “subfigure” zooms over the first rounds, to provide a closer
look to the test cases that converge quickly. The right part of
Figure~\ref{fig:plot-avg-max} shows the maximum error (computed over all nodes,
and over $50$ experiments) for all our graphs (points have been slightly
translated to improve visualization). As it can be seen, in all our
experimental data sets, the maximum error is at most equal to 1 by cycle 22.

These error figures tell us that if the exact computation of coreness is not required
(for example if coreness is used to optimize gossip protocols
in a social network), the $k$-core decomposition algorithms proposed 
may be stopped after a predefined number of rounds, knowing that
both the average and the maximum errors would be extremely low.

\subsection{One-to-many version}

The main reason for running the one-to-many version of the protocol is to 
compute the $k$-core decomposition over large graphs, that cannot
fit into the memory of a single machine. Experimental results showed
that the number of rounds needed to complete the protocol was
equivalent to that of the one-to-one version. 
One of the key performance figures to be considered for the
one-to-many version is
the communication overhead generated by update messages exchanged among
hosts. The overhead is computed as the average number of times
a node generates a new estimate that has to be sent to another host.
 
\begin{figure}
\begin{center}
\includegraphics[width=9cm]{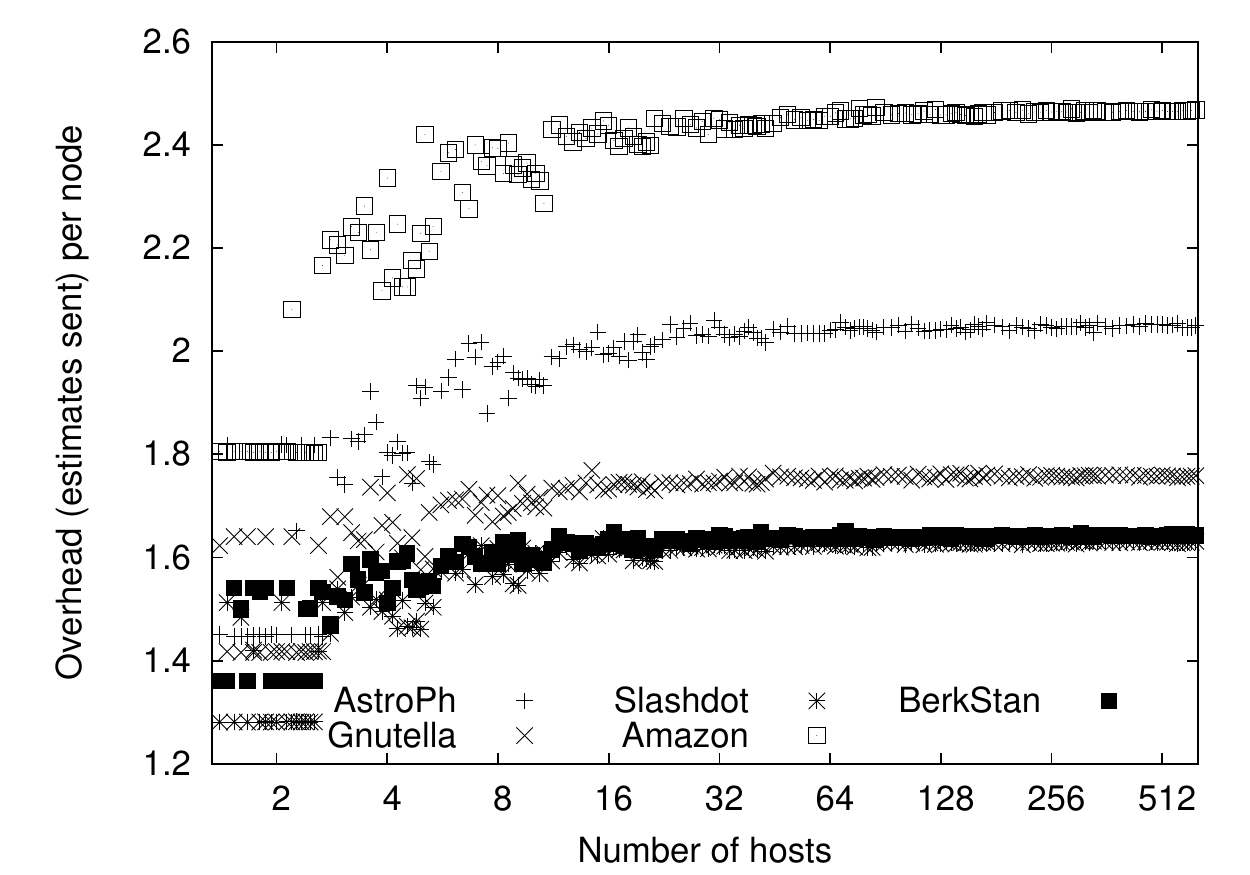}
\includegraphics[width=9cm]{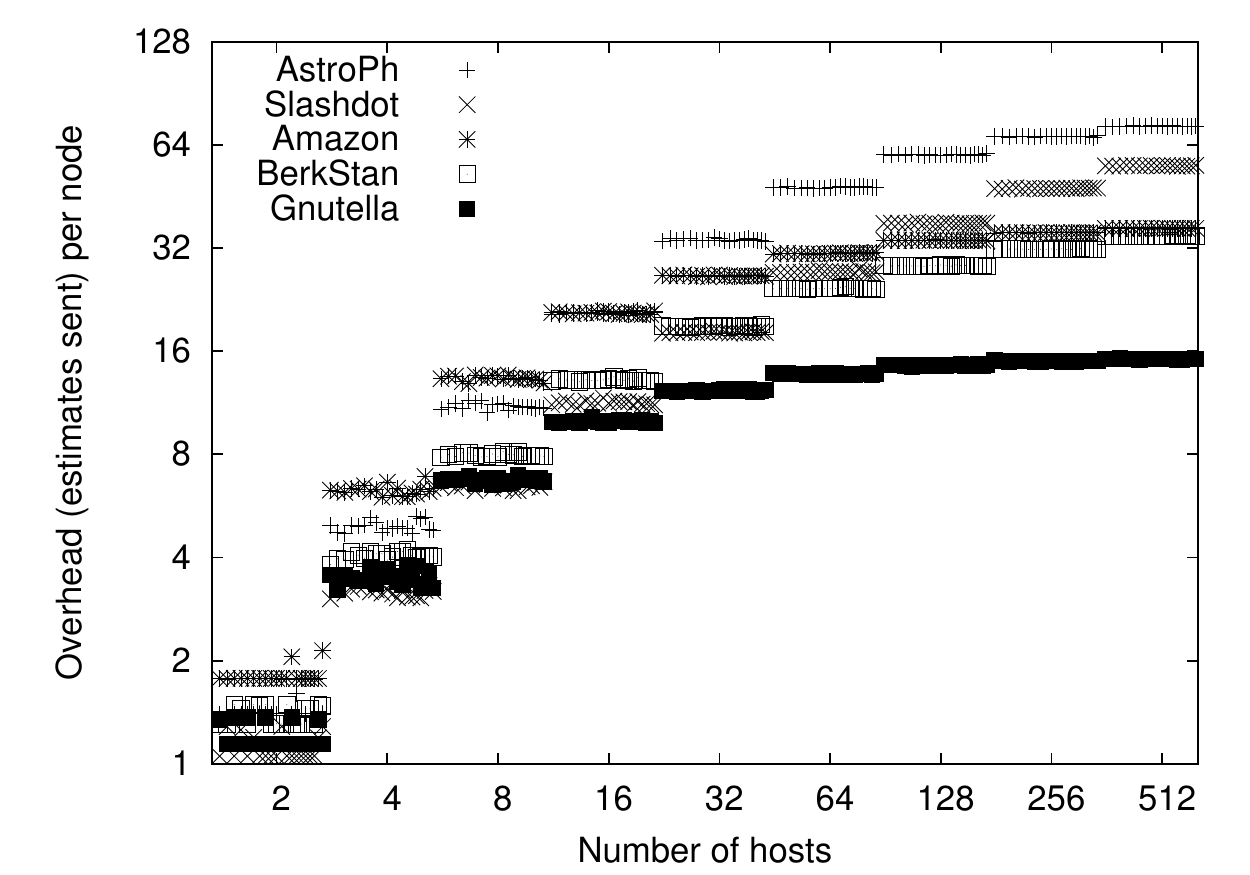}
\end{center}
\caption{Overhead per node -- with (left) and without (right) broadcast medium.}
\label{fig:cluster-size}
\end{figure}

Figure~\ref{fig:cluster-size} shows the overhead per node with a variable
number of hosts, with (left) and without (right) a medium broadcast available.
For visualization reasons, only some of the original data sets have been
considered; but the results are similar for all of them. Twenty
experiments were considered for this case. In the graph, the outcome
of each experiment was represented as a point (slightly translated
for the sake of visualization clarity).

When a broadcast medium is not available and point-to-point communication is used, the overhead increases with the
number of hosts available, tending to stabilize to the levels of the one-to-one
protocol (see the $m_\textit{avg}$ column of Table~\ref{fig:table}
--- values are slightly higher given that the optimization of Section~\ref{s:optim}
cannot be applied in this case). When a broadcast medium is available, on the
other hand, the efficiency is much higher. In this case, a single message
is sent at each round, containing all the estimates that have changed since
the previous one. Most of the nodes reach the correct estimate after 
few rounds and very few estimates are sent on their behalf after the first rounds;
the effect is that the average number of estimates sent per node is extremely
low, always smaller than $3$, making the one-to-many algorithm particularly 
well-suited for clusters connected through fast local area networks.

\section{Conclusions}

To the best of our knowledge, this paper is the first to propose distributed
algorithms for the $k$-core decomposition of online and/or large graphs. While
theoretical analysis provided us with fairly large upper bounds on the number
of rounds required to complete the algorithm, which are strict for specific
worst-case examples, experimental results have shown that for realistic graphs,
our algorithms efficiently converge in few rounds.

The next logical step is the actual implementation of the algorithms. For this
purpose, we are considering distributed frameworks like
Hadoop~\cite{map-reduce} and Pregel~\cite{pregel}, in which the computation is
divided in logical units (corresponding to the collection of nodes under the
responsibility of a single host) and these units are divided among a
collection of computational processes, termed workers, in charge of
processing them according to a set of defined rules. 
This would allow our solutions to inherit the desirable features of
these frameworks in terms of efficiency, scalability and fault
tolerance.
\bibliographystyle{acm}
\bibliography{references}

\end{document}